\documentclass[aps,prl,superscriptaddress,reprint]{revtex4-1}
\usepackage{amsmath}
\usepackage{amssymb}
\usepackage{graphicx}
\usepackage[colorlinks,urlcolor=blue]{hyperref}
\usepackage{url}
\usepackage{color,xcolor}
\usepackage{ulem}
\usepackage{float}
\usepackage{epstopdf}
\usepackage{changes}
\begin{document}

\title{Prediction of $s^\pm$-wave superconductivity enhanced by electronic doping in trilayer nickelates La$_4$Ni$_3$O$_{10}$ under pressure}
\author{Yang Zhang}
\author{Ling-Fang Lin}
\affiliation{Department of Physics and Astronomy, University of Tennessee, Knoxville, Tennessee 37996, USA}
\author{Adriana Moreo}
\affiliation{Department of Physics and Astronomy, University of Tennessee, Knoxville, Tennessee 37996, USA}
\affiliation{Materials Science and Technology Division, Oak Ridge National Laboratory, Oak Ridge, Tennessee 37831, USA}
\author{Thomas A. Maier}
\affiliation{Computational Sciences and Engineering Division, Oak Ridge National Laboratory, Oak Ridge, Tennessee 37831, USA}
\author{Elbio Dagotto}
\affiliation{Department of Physics and Astronomy, University of Tennessee, Knoxville, Tennessee 37996, USA}
\affiliation{Materials Science and Technology Division, Oak Ridge National Laboratory, Oak Ridge, Tennessee 37831, USA}

\date{\today}

\begin{abstract}
Motivated by the recently reported signatures of superconductivity in trilayer La$_4$Ni$_3$O$_{10}$ under pressure, we comprehensively study this system using {\it ab initio} and random-phase approximation techniques. Without electronic interactions, the Ni $d_{3z^2-r^2}$ orbitals show a bonding-antibonding and nonbonding splitting behavior via the O $p_z$ orbitals inducing a ``trimer'' lattice in La$_4$Ni$_3$O$_{10}$, analogous to the dimers of La$_3$Ni$_2$O$_{7}$. The Fermi surface consists of three electron sheets with mixed $e_g$ orbitals, and a hole and an electron pocket made up of the $d_{3z^2-r^2}$ orbital, suggesting a Ni two-orbital minimum model. In addition, we find that superconducting pairing is induced in the $s^{\pm}$-wave channel due to partial nesting between the {\bf M}=$(\pi, \pi)$ centered pockets and portions of the Fermi surface centered at the {\bf $\Gamma$}=$(0, 0)$ point. With changing electronic density $n$, the $s^\pm$ instability remains leading and its pairing strength shows a dome-like behavior with a maximum around $n = 4.2$ ($\sim 6.7\%$ electron doping). The superconducting instability disappears at the same electronic density as that in the new 1313 stacking La$_3$Ni$_2$O$_7$, correlated with the vanishing of the hole pocket that arises from the trilayer sublattice, suggesting that the high-$T_c$ superconductivity of La$_3$Ni$_2$O$_7$ does $not$ originate from a trilayer- and single-layer structure. Furthermore, we confirm the experimentally proposed spin state in La$_4$Ni$_3$O$_{10}$ with an in-plane ($\pi$, $\pi$) order and antiferromagnetic coupling between the top and bottom Ni layers, and spin zero in the middle layer.
\end{abstract}

\maketitle
{\it Introduction.--} The discovery of superconductivity in the bilayer Ruddlesden-Popper (RP) perovskite La$_3$Ni$_2$O$_7$ (327-LNO) with a $d^{\rm 7.5}$ configuration under high pressure~\cite{Sun:arxiv} opened a remarkable platform for the study of nickelate-based superconductors~\cite{LiuZhe:arxiv,Zhang:arxiv-exp,Hou:arxiv,Yang:arxiv09,Zhang:arxiv09,Wang:arxiv9,Luo:arxiv,Zhang:arxiv,Christiansson:arxiv,
Yang:arxiv,Sakakibara:arxiv,Shen:arxiv,Liu:arxiv,Zhang:arxiv1,Yang:arxiv1,Oh:arxiv,Liao:arxiv,Cao:arxiv,Lechermann:arxiv,Shilenko:arxiv,Jiang:arxiv,
Huang:arxiv,Zhang:prb23,Qin:arxiv,Zhang:prb24}. By increasing pressure, 327-LNO transforms from the Amam to the Fmmm structure, the latter without tilting of oxygen octahedra~\cite{Sun:arxiv}. Superconductivity was reported in a broad pressure range from 14 to 43.5 Gpa in the Fmmm phase, with the transition temperature $T_c$ $\sim 80$ K~\cite{Sun:arxiv}.

To explore superconductivity in other RP layered nickelates, both theoretical and experimental studies have expanded to single-layer La$_2$NiO$_4$ and trilayer (TL) La$_4$Ni$_3$O$_{10}$ (4310-LNO) systems~\cite{Li:nc,Puggioni:prb,Zhang:nc20,Zhang:prm,Rout:prb,Segedin:nc,Zhang:arxiv09,Yuan:arxiv11,Leonov:arxiv12,Samarakoon:prx}, but no superconductivity was found at ambient conditions.  Interestingly, based on neutron diffraction extinction rules, a novel magnetic state with up-zero-down in trilayers was inferred by J. Zhang et al.~\cite{Zhang:nc20} and our results confirm those predictions. In addition, superconductivity was absent also in La$_2$NiO$_4$ under pressure~\cite{Zhang:arxiv09}.

\begin{figure*}
\centering
\includegraphics[width=0.80\textwidth]{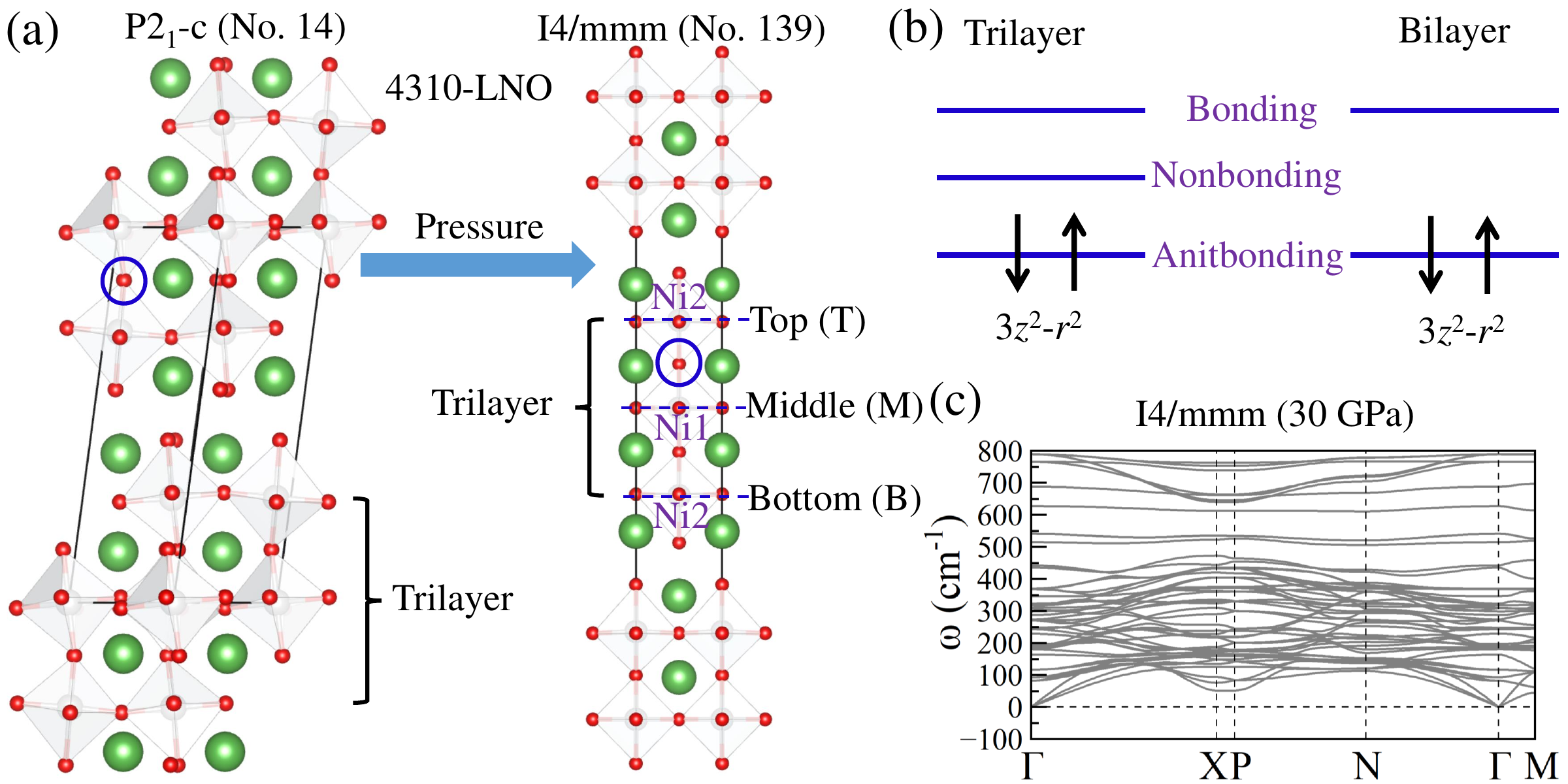}
\caption{(a) Schematic crystal structure of the conventional cells of TL 4310-LNO for the P2$_1$-c and I4/mmm phases without pressure and at high pressure, respectively (green = La; gray = Ni; red = O). Note that the local $z$-axis is perpendicular to the NiO$_6$ plane towards the top O atom, while the local $x$- or $y$-axis are along the in-plane Ni-O bond directions. All crystal structures were visualized using the VESTA code~\cite{Momma:vesta}. (b) Sketches of the $d_{3z^2-r^2}$ orbital in the TL and bilayer nickelates with two electrons. (c) Phonon spectrum of the I4/mmm phase of TL 4310-LNO at 30 GPa.}
\label{Crystal}
\end{figure*}

However, very recently, signatures of superconductivity were also reported in another RP perovskite nickelate La$_4$Ni$_3$O$_{10}$ (4310-LNO), with $T_c$ about $20-30$ K above 15 GPa~\cite{Sakakibara:arxiv09,Li:cpl,Zhu:arxiv11,Zhang:arxiv11,Li:arxiv11,Wang:arxiv12}. Without pressure, 4310-LNO has a monoclinic P2$_1$-c structure (No. 14)~\cite{Puggioni:prb,Yuan:arxiv11}, where the strongly distorted corner-sharing NiO$_6$ octahedra form a TL sublattice stacking along the $c$-axis (see Fig.~\ref{Crystal}(a)). Under the influence of hydrostatic pressure, 4310-LNO also shows a structural phase transition from P2$_1$-c symmetry to a high-symmetry I4/mmm phase without the tilting of oxygen octahedra, similarly to 327-LNO~\cite{Zhu:arxiv11}.

Thus, considering these developments on 327-LNO and 4310-LNO, several interesting questions naturally arise: What are the similarities and differences between the bilayer 327-LNO and TL 4310-LNO nickelates under pressure? What is the superconducting pairing channel in 4310-LNO? How does superconductivity in 4310-LNO evolve under electron doping?

{\it Trimer vs dimer --} Similar to the ``dimer'' physics in the bilayer lattice~\cite{Zhang:arxiv,Dagotto:prb1992}, the ``trimer'' physics can also be obtained in the TL lattice because the intraorbital coupling is strong and the coupling in between TLs is weak. Specifically, the $d_{3z^2-r^2}$ orbital would split into antibonding, nonbonding, and bonding states in the TL 4310-LNO, as shown in Fig.~\ref{Crystal}(b). Because the $d_{x^2-y^2}$ orbital is lying in the NiO$_6$ plane, it remains decoupled among planes, not participating in the formation of the antibonding-nonbonding splitting along the $z$-axis, resulting in an orbital-selective behavior~\cite{Streltsovt:prb14,Zhang:ossp}.

In 4310-LNO, the total electronic density of Ni is $d^{7.33}$, corresponding to Ni$^{2.67}$ on average, leading to partially filled $e_g$ orbitals and three fully occupied $t_{2g}$ states. In this case, the $d_{3z^2-r^2}$ orbital is nearly half-filled, and the $d_{x^2-y^2}$ orbital is close to one-third occupied.  Due to large in-plane interorbital hopping between the $e_g$ states, the ``self-doped''  behavior of the $e_g$ orbitals is also expected in the TL 4310-LNO, similar to 327-LNO~\cite{Zhang:arxiv1}.

To better understand these broad issues, using first-principles density functional theory (DFT)~\cite{Kresse:Prb,Kresse:Prb96,Blochl:Prb,Perdew:Prl}, we have studied the TL 4310-LNO in detail. Without pressure, our DFT results find that the P2$_1$-c phase has an energy lower by about -48.26 meV/Ni than the I4/mmm phase. By introducing  pressure, the monoclinic distortion is gradually suppressed, leading to a high-symmetry I4/mmm phase at high pressure, in agreement with previous experimental works~\cite{Zhu:arxiv11,Li:arxiv11}. Furthermore, the phononic calculations indicate that the I4/mmm phase of 4310-LNO is stable without any imaginary frequency at 30 GPa~\cite{HP}, as displayed in Fig.~\ref{Crystal}(c), by using the density functional perturbation theory approach~\cite{Baroni:Prl,Gonze:Pra1}, analyzed by the PHONONPY software~\cite{Chaput:prb,Togo:sm}. Thus, the pressure effect is quite similar in 4310-LNO and 327-LNO~\cite{Zhang:arxiv1}, where the spontaneous suppression of octahedral distortion under pressure leads to a phase transition from low to high symmetry.

{\it Electronic structures of LNO --} Near the Fermi level, the main contributions to the electronic density of states are from the Ni $3d$ orbitals hybridized with the O $p$ orbitals with a large charge-transfer energy $\Delta$ = $\varepsilon_{d}$ - $\varepsilon_{p}$, sharing the common character of other nickelates~\cite{Zhang:prb20,Nomura:rpp}. Using the maximally localized Wannier functions~\cite{Mostofi:cpc} by fitting DFT and Wannier bands of the non-magnetic state of the I4/mmm phase of 4310-LNO at 30 GPa, we find that both $e_g$ orbitals of the outer layer Ni have lower onsite energies than that in inner layer Ni. The nearest-neighbor (NN) hopping of the $d_{3z^2-r^2}$  orbital along the $z$-axis for 4310-LNO ($\sim 0.694$ eV) is slightly larger than that in 327-LNO ($\sim 0.640$ eV)~\cite{Luo:arxiv,Zhang:arxiv}. In the Ni plane, the largest hopping is the intraorbital hopping of the $d_{3x^2-y^2}$  orbital ($\sim 0.519/0.511$ eV for inner and outer layer Ni). Furthermore, we also obtain a large {\it interorbital} hopping between $d_{3z^2-r^2}$ and $d_{x^2-y^2}$ orbitals in 4310-LNO, caused by the ligand ``bridge'' of the in-plane O $p_x$ or $p_y$ orbitals connecting those two orbitals.

Next, we constructed a six-band ${e_g}$-orbital tight binding (TB) model on the TL lattice for the I4/mmm phase of 4310-LNO at 30 GPa with overall filling $n = 4$ by using the NN and next nearest-neighbor (NNN) hoppings, similar to another independent  work~\cite{Luo:arxiv24}. More details can be found in Supplementary Material~\cite{Supplemental}.

\begin{figure}
\centering
\includegraphics[width=0.42\textwidth]{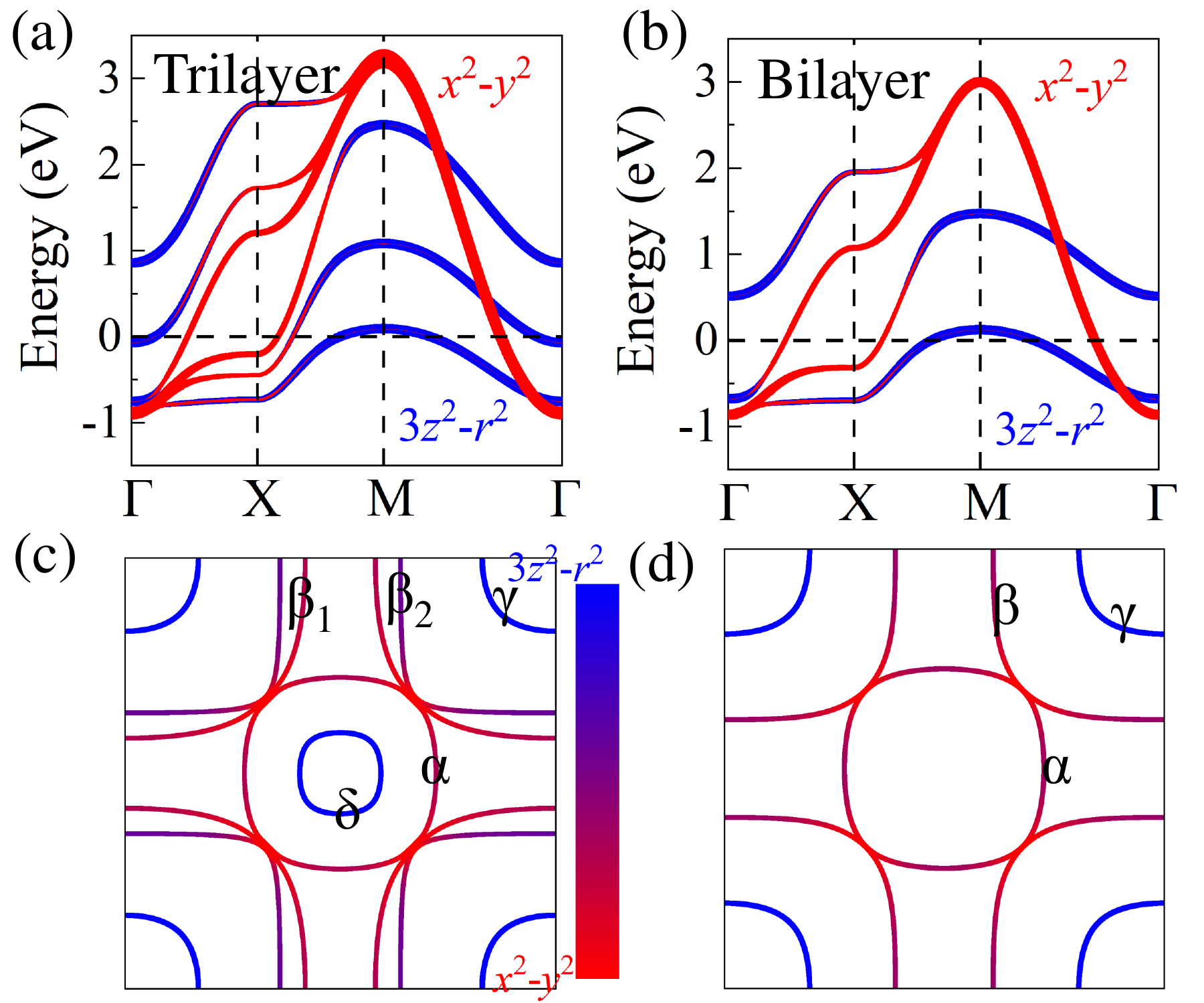}
\caption{ TB band structures and FS's for (a,c) TL 4310-LNO, and (b,d) bilayer 327-LNO,  respectively. (a,c) The six-band $e_g$ orbital TB model was considered with three NN and NNN hoppings in a TL lattice for the overall filling $n$ = 4 ($4/3$ electrons per site). (b,d) The four-band $e_g$ orbital TB model was considered in a bilayer lattice for the overall filling $n$ = 3 (1.5 electrons per site), where the hoppings used from a previous study~\cite{Luo:arxiv}.}
\label{TB}
\end{figure}

As shown in Fig.~\ref{TB}(a), $d_{3z^2-r^2}$ displays the bonding-antibonding, and nonbonding splitting behavior, while the $d_{x^2-y^2}$ orbital remains decoupled among planes, in agreement with our discussion in the previous section. Compared with the bilayer 327-LNO (see Fig.~\ref{TB}(b)), the bandwidth of the $e_g$ orbitals increases by about $\sim 8 \%$. The calculated avenge electronic densities are 2.085 and 1.915 for the $d_{3z^2-r^2}$ and $d_{x^2-y^2}$ orbitals (0.695 and 0.638 per site), respectively, in the TL TB model of 4310-LNO, while they are 1.682 and 1.318 for 327-LNO (0.814 and 0.659 per site). Considering the average valences of the Ni ions in 4310-LNO and 327-LNO, holes are favored to enter the $d_{3z^2-r^2}$ orbitals.

Five bands are crossing the Fermi level in 4310-LNO at high pressure, contributing to the Fermi surface (FS) as displayed in Fig.~\ref{TB}(c), namely, bands $\alpha$, $\beta_1$ $\beta_2$, $\gamma$ and $\delta$, respectively. Similarly to the FS of 327-LNO (see Fig.~\ref{TB}(d)), the hole pocket $\gamma$ is made up by the $d_{3z^2-r^2}$ orbital, while the three electron sheets $\alpha$, $\beta_1$ and $\beta_2$ originate from mixed $d_{3z^2-r^2}$ and $d_{x^2-y^2}$ orbitals. In addition, an electron pocket $\delta$ made up by the nonbonding $d_{3z^2-r^2}$ orbital is obtained for 4310-LNO.

{\it RPA pairing tendencies --} Next, we have used multi-orbital random phase approximation (RPA) calculations to assess the bilayer TB models for their superconducting behavior. The RPA is based on a perturbative weak-coupling expansion in the Coulomb interaction~\cite{Kubo2007,Graser2009,Altmeyer2016,Romer2020}. The pairing strength $\lambda_\alpha$ for channel $\alpha$  and the corresponding gap structure $g_\alpha({\bf k})$ are obtained by solving an eigenvalue problem of the form
\begin{eqnarray}\label{eq:pp}
	\int_{FS} d{\bf k'} \, \Gamma({\bf k -  k'}) g_\alpha({\bf k'}) = \lambda_\alpha g_\alpha({\bf k})\,,
\end{eqnarray}
where the momenta ${\bf k}$ and ${\bf k'}$ are on the FS, and $\Gamma({\bf k - k'})$ is the irreducible particle-particle vertex. In the RPA approximation, the dominant term entering $\Gamma({\bf k-k'})$ is the RPA spin susceptibility $\chi({\bf k-k'})$.

\begin{figure}
\centering
\includegraphics[width=0.42\textwidth]{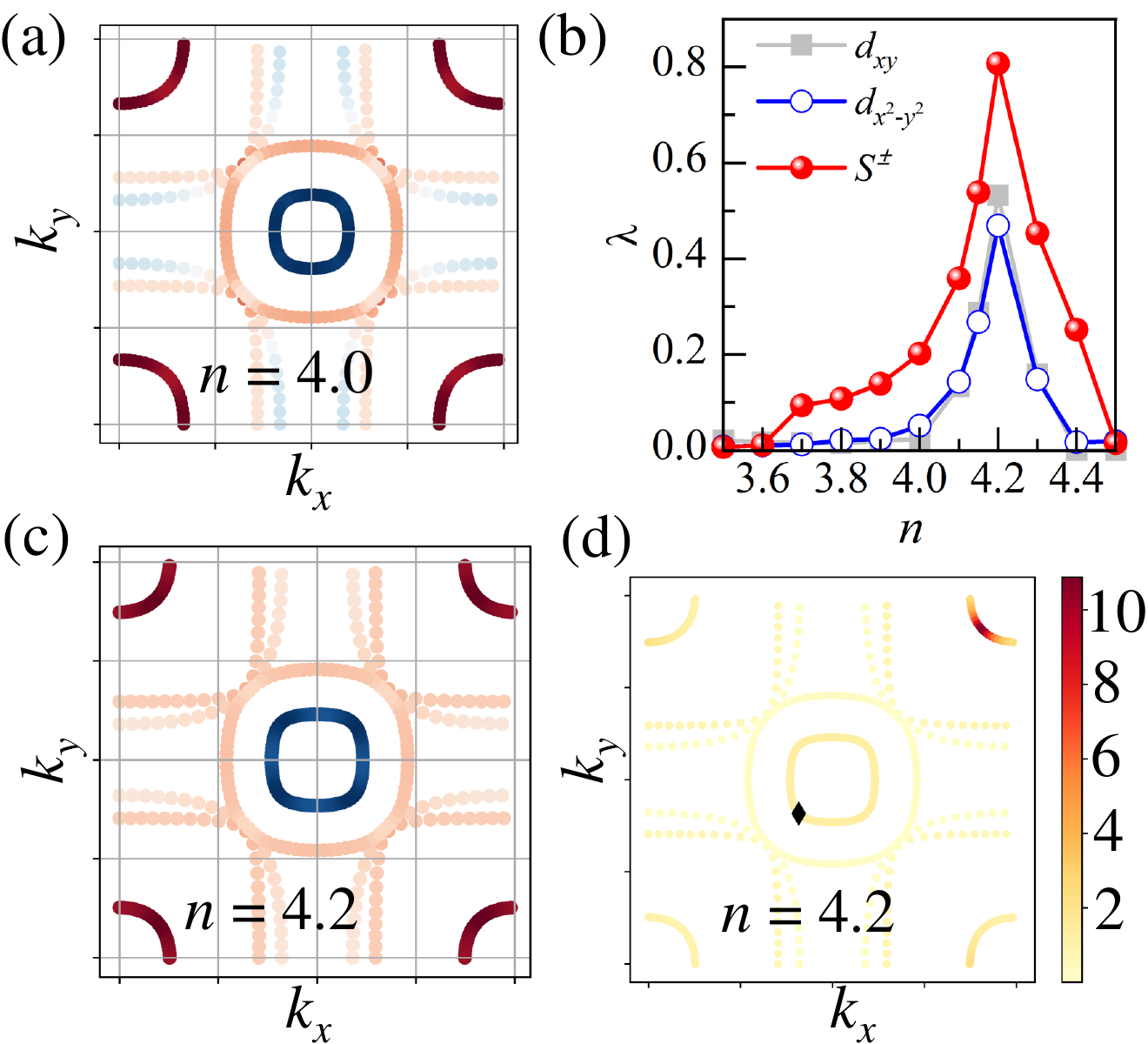}
\caption{(a) The calculated RPA superconducting gap structure $g_\alpha({\bf k})$ for momenta ${\bf k}$ on the FS of 4310-LNO with $s_\pm$-wave symmetry at $n = 4.0$. The sign of the gap is indicated by the colors (red = positive, blue = negative), and the gap amplitude by its intensity. (b) The RPA calculated pairing strength $\lambda$ for the $s^\pm$, $d_{x^2-y^2}$ and $d_{xy}$ instabilities versus electronic densities for the TL model. (c-d) The calculated RPA (c) superconducting gap structure $g_\alpha({\bf k})$ with $s_\pm$-wave symmetry and (d) the pairing interaction $\Gamma$(${\bf k}$, ${\bf k}_0$) with ${\bf k}_0$ indicated by the black diamond for $n = 4.2$. The RPA calculations used $U = 0.95$, $U'=U/2$ and $J = J' = U/4$ in units of eV ($J$ is the Hund coupling, $J'$ the pair hopping), with NN and NNN hoppings from the I4/mmm phase of 4310-LNO. }
\label{Pairing}
\end{figure}

By solving the eigenvalue problem in Eq.~(\ref{eq:pp}) for the RPA pairing interaction of 4310-LNO ($n = 4.0$), we find that the $s^\pm$ gap structure is the leading pairing symmetry caused by spin fluctuations. The gap is large and switches sign between the small electron pocket at $\Gamma$ and the small hole pocket at $M$, which are separated by ($\pi$, $\pi$) (see Fig.~\ref{Pairing}(a)). The calculated pairing strength $\lambda$ ($\sim 0.202$) of the $s^\pm$-wave gap structure is smaller than that in 327-LNO at the same $U = 0.95$ ($\sim 0.39$). Note that superconductivity was also predicted by an earlier recent study but they did not discuss the pairing channel~\cite{Sakakibara:arxiv09}. Since in our RPA treatment, the pairing strength $\lambda$ enters exponentially in the equation for $T_c$, i.e. $T_c=\omega_0e^{-1/\lambda}$ with a spin-fluctuation cut-off frequency $\omega_0$, this comparison suggests a lower $T_c$ for 4310-LNO than 327-LNO to the extent that $\omega_0$ is similar in both systems. Although the $s^\pm$-wave superconducting pairing symmetry was also obtained in bilayer 327-LNO~\cite{Zhang:arxiv1}, the nesting vector is different from 4310-LNO. Independently, another work also proposed the same $s^\pm$-wave and a lower $T_c$ in 4310-LNO~\cite{Yang:arxiv24}.

To understand doping effects, we also studied the dependence of the RPA pairing strength $\lambda$ on the electron density $n$ in the TL model, as shown in Fig.~\ref{Pairing}(b). One sees that the $s^\pm$ state remains leading over the entire density range we have studied, while $d_{x^2-y^2}$ and $d_{xy}$ states are subleading. Remarkably, the pairing strengths $\lambda$ for all three states show a dome-like doping dependence with a peak at $n = 4.2$ ($\sim 6.7\%$ electron doping). Near $n = 3.6$ ($\sim 13.3\%$ hole doping) or $n = 4.5$ ($\sim 16.7\%$ electron doping), the calculated RPA pairing strength $\lambda$ becomes negligible, indicating that a superconducting instability may be absent beyond the central dome. The leading $s^\pm$ gap for the optimal density $n=4.2$ is shown in Fig.~\ref{Pairing}(c), and the corresponding pairing interaction $\Gamma({\bf k-k}_0)$ for this case in Fig.~\ref{Pairing}(d). Here, ${\bf k}_0$ is fixed at the Fermi momentum on the inner $\Gamma$-centerd pocket indicated by the black diamond, and ${\bf k}$ runs along all the Fermi surface points. We see that $\Gamma({\bf k-k}_0)$ is large and peaked for a momentum transfer of ${\bf q} \sim (\pi,\pi)$ that connects states on the inner $\Gamma$ pocket and on the $M$-centered pocket. This pair scattering drives the leading $s^\pm$ state which has a large gap with opposite signs on these Fermi surface sheets, as seen in Fig.~\ref{Pairing}(c).

\begin{figure*}
\centering
\includegraphics[width=0.94\textwidth]{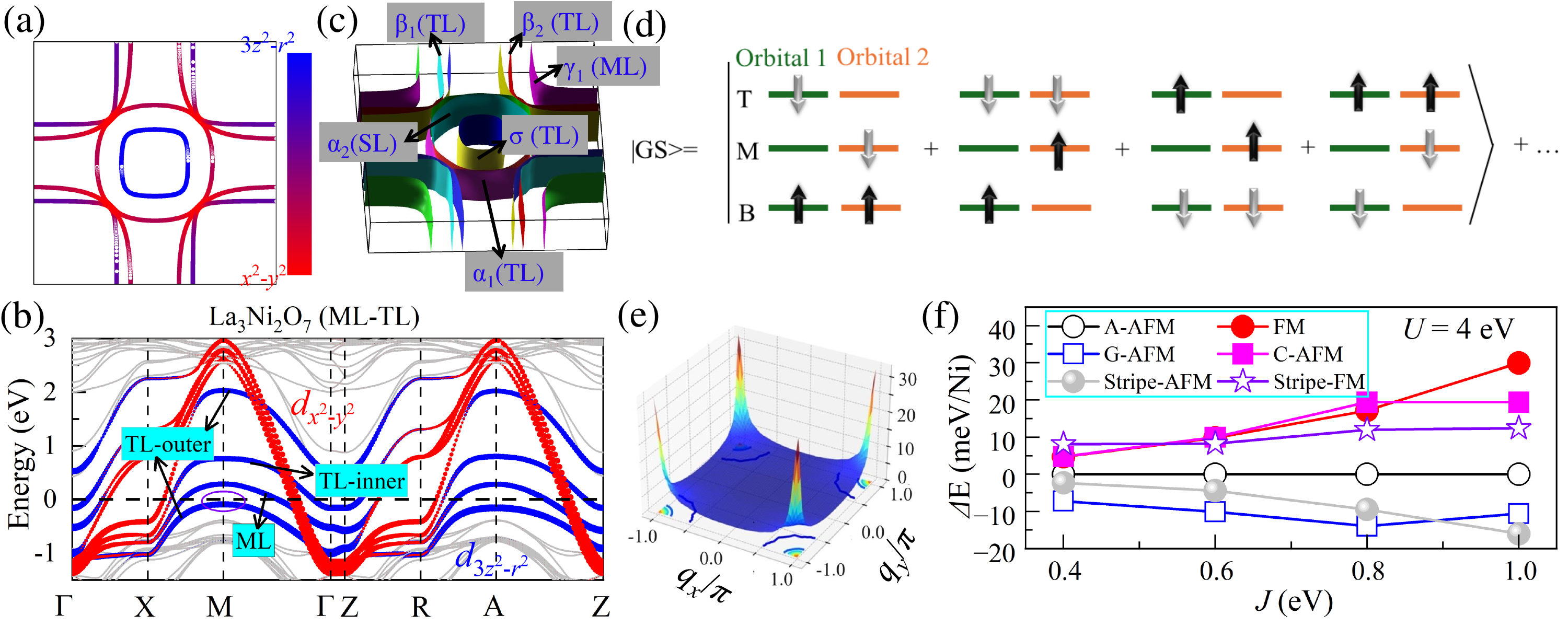}
\caption{(a) TB FS for $n = 4.5$ with the absence of the $\gamma$ pocket. (b) The DFT band structure and (c) calculated FS for La$_3$Ni$_2$O$_7$ with alternating ML and TL structures using the experimental structure at high pressure. (d) The state with the largest coefficient in absolute value in the ground state of the exactly diagonalized three-Ni-site TL model, along the direction perpendicular to the layers. (e) The RPA calculated static spin susceptibility $\chi'({\bf q}, \omega=0)$ versus $q_x$, $q_y$ for $q_z=\pi/2$ for the two-orbital TL TB model for $n=4.2$. (f) The DFT+$U$+$J$ calculated energies for different $J$'s of different magnetic configurations at $U  = 4$ eV.}
\label{RPA}
\end{figure*}

In addition, we find that the superconducting pairing strength at $n = 4.5$, corresponding to 1.5 electrons per site, is almost zero. For this case, we find that the hole $\gamma$ pocket is absent, as shown in Fig.~\ref{RPA}(a). Futhermore, very recently, several groups independently reported a new phase of La$_3$Ni$_2$O$_7$ with alternating monolayer (ML) and TL structures~\cite{Chen:arxiv12,Puphal:arxiv12,Xu:arxiv12}. Note that the electronic density of the $e_g$ orbitals of La$_3$Ni$_2$O$_7$ is also 1.5 per Ni. We therefore calculated the band structure of the $P4/mmm$ phase of this new phase of La$_3$Ni$_2$O$_7$ by using the experimental structure under high pressure~\cite{LNO}. Figure~\ref{RPA}(b) indicates that the $\gamma$ pocket of the $d_{3z^2-r^2}$ orbital contributed by the TL structure is absent in this new phase of La$_3$Ni$_2$O$_7$. As shown in Fig.~\ref{RPA}(c), the two $\beta$ sheets, $\sigma$ pocket, and $\alpha_1$ sheet are mainly induced by the TL sublattice in the alternating stacking La$_3$Ni$_2$O$_7$, while the $\alpha_2$ and $\gamma_1$ sheets arise from the ML sublattice. Very recent experiments also suggest the hole $\gamma$ pocket contributed by the TL sublattice is absent in this new phase of La$_3$Ni$_2$O$_7$ without pressure~\cite{Abadi:arxiv24}.

Our RPA results show that the superconductivity instability disappears at $n = 4.5$ correlated with the absence of the $\gamma$ pocket in the TL lattice model. For the ML in the alternating stacking La$_3$Ni$_2$O$_7$, it is possible to obtain the superconducting instability. Considering the alternating ML and TL stacking, the superconductivity would be likely restricted to the ML where the effective coupling between two MLs should be quite weak because the TL is between two MLs. Therefore, the Kosterlitz-Thouless order will not be converted to regular long-range order, or $T_c$ should not be high if long-range order can be established. However, a robust superconductivity instability was obtained for the bilayer 327-LNO with same electronic density~\cite{Zhang:arxiv1}. Thus, our results suggest that the previously discovered high-$T_c$ superconductivity in La$_3$Ni$_2$O$_7$ does not originate from an alternating ML and TL stacking structure~\cite{1313}.

{\it Magnetic tendency --} To understand intuitively the magnetic tendency in 4310-LNO, we first diagonalized the two-orbital TL model with $U$, $U^{'}$, $J_H$,  and the hoppings and crystal field for a small cluster with 3 sites, only along the $z$-axis. Then, we obtained the exact-diagonalization ground state of this cluster. The dominant state (largest coefficient) in the ground state is displayed in Fig.~\ref{RPA}(d), where the top and bottom layers are antiferromagnetically (AFM) coupled. In addition, the density is larger in the top and bottom layers than in the middle layer.

We also studied the static RPA enhanced spin susceptibility $\chi'({\bf q}, \omega=0)$ that is obtained from the Lindhart function $\chi_0({\bf q})$ as
\begin{eqnarray}
\chi({\bf q}) = \chi_0({\bf q})[1-{\cal U}\chi_0({\bf q})]^{-1}.
\end{eqnarray}
$\chi_0({\bf q})$ is an orbital-dependent susceptibility tensor and ${\cal U}$ is a tensor involving the interaction parameters \cite{Graser2009}.

$\chi({\bf q})$ for $n=4.2$ presents a strong peak at ${\bf q}$ = ($\pi$, $\pi$, $\pi/2$), as displayed in Fig.~\ref{RPA}(e). This spin density wave fluctuation corresponds to a G-type AFM state in which the top and bottom layers are coupled antiferromagnetically both in-plane and between the planes, as in the 3-sites cluster results of panel Fig.~\ref{RPA}(d) and where the middle layer has zero spin density.

To confirm the RPA results, we also studied the magnetic properties by using the DFT+$U$+$J$ formalism within the Liechtenstein formulation with a double-counting term to deal with the onsite Coulomb interactions~\cite{Liechtenstein:prb}, where $U$ is fixed at 4 eV, following recent DFT studies of nickelates~\cite{Zhang:arxiv1,Christiansson:arxiv}. Here, we considered several possible magnetic structures of the Ni TL spins with spin-zero in the middle layer as input:  (1) A-AFM or ferromagnetic (FM) with in-plane wavevector (0, 0) where top and bottom are AFM or FM coupled; (2) G-AFM or C-AFM with in-plane wavevector ($\pi$, $\pi$) where top and bottom layers are AFM or FM coupled; (3) Stripe-AFM or Stripe-FM with in-plane wavevector ($\pi$, 0), where the top and bottom layers are AFM or FM coupled.

As displayed in Fig.~\ref{RPA}(f), the G-AFM state has the lowest energy when $J \textless 1$ eV among all the candidates. In addition, we also considered the cases with nonzero spin in the middle layer. Those spin states were found to have higher energy than the cases with spin zero in the middle layer. Considering the previously calculated $J$ for other layered nickelates ($\sim 0.61 - 0.68$ eV)~\cite{Christiansson:arxiv,Botana:prb,Pardo:prb}, our DFT+$U$+$J$ calculations also found the in-plane ($\pi$, $\pi$) order with AFM coupling between top and bottom Ni layers, while the middle layer has spin zero, in agreement with the RPA calculations. Thus, our theoretical results corroborate the up-zero-down picture previously inferred from experiments~\cite{Zhang:nc20,Samarakoon:prx}.

{\it Conclusions.--} In summary, we have unveiled clear similarities and differences between the TL nickelate and the recently much-discussed bilayer 327-LNO nickelates. (1) Similar to 327-LNO, pressure spontaneously suppresses the octahedral distortion in the TL, leading to a phase transition from a low- to a high-symmetry phase in 4310-LNO as well as to a large in-plane interorbital hopping between the $e_g$ states. (2) The Ni $d_{3z^2-r^2}$ orbital shows a bonding-antibonding splitting, but also has a nonbonding state in 4310-LNO due to the geometry of the Ni TL lattice. (3) The 4310-LNO Fermi surface contains three electron sheets formed by mixed $e_g$ orbitals, and a hole and an electron pocket made of the $d_{3z^2-r^2}$ orbital, establishing that a minimum two $e_g$ orbital model per Ni is needed. (4) We also found a leading spin-fluctuation driven $s^\pm$-wave pairing state in 4310-LNO, where the gap is largest and has opposite signs on the small electron pocket at $\Gamma$ and the small hole pocket at $M$, which are separated by ($\pi$, $\pi$). (5) Under variation of the electron density $n$, the pairing strength displays dome-like behavior and is strongly enhanced for $n=4.2$($\sim 6.7\%$ electron doping) before it becomes negligibly weak at $n = 4.5$, correlated with the disappearance of the $M$-centered $\gamma$ pocket. (6) We also discussed the interesting spin density wave state with in-plane ($\pi$, $\pi$) spin order, zero spin density in the middle layer, and AFM coupling between the top and bottom layers in 4310-LNO.

This work was supported by the U.S. Department of Energy (DOE), Office of Science, Basic Energy Sciences (BES), Materials Sciences and Engineering Division.


\begin{references}
\bibitem{HP}{In our calculations, the monoclinic distortion is already fully suppressed, leading to the high-symmetry I4/mmm phase at 30 GPa. Thus,  we will focus on the results of the I4/mmm phase of 4310-LNO. The detailed phase transition under pressure is not the scope of our present work, thus we leave this issue to future work.}
\bibitem{LNO}{Here, based on the high-pressure lattice structure~\cite{Puphal:arxiv12}, we fully relaxed the atomic positions until the Hellman-Feynman force on each atom was smaller than $0.001$ eV/{\AA}.}
\end{references}
\end{document}